# Melt–Quench Failures and Practical Solutions for Universal Machine-Learning Interatomic Potentials in Amorphous Structure Generation

Shuwei Li[1†], Yuqi An[1†], Xingyu Guo[2], Wenqiang Yang[3*], Zhenbin Wang[1,4*]

[1]Department of Materials Science and Engineering, City University of Hong Kong, Hong Kong SAR, 999077, China

[2]Department of Data Science, City University of Hong Kong, Hong Kong SAR, 999077, China

[3]Department of Chemical Engineering, University of South Carolina, Columbia, South Carolina, USA

[4]School of Energy and Environment, City University of Hong Kong, Hong Kong SAR, 999077, China

[†]These authors contributed equally to this work.

*Corresponding author, E-mail: wenqiang@email.sc.edu; zwan22@cityu.edu.hk

## Abstract

Generating experimentally relevant amorphous structures via melt–quench molecular dynamics is prohibitively expensive at the first-principles level. Universal machine-learning interatomic potentials (uMLIPs) could accelerate such simulations, but their reliability under non-equilibrium conditions remains unclear. Here, we examine eight leading uMLIPs for generating amorphous $IrO_2$, using this electrocatalytically relevant oxide as a diagnostic case. Under the conventional melt–quench protocol, all models yield unphysically expanded structures with densities of 1–4 g/cm$^3$, far below the ab initio molecular dynamics (AIMD) reference value of 10.04 g/cm$^3$. Comparisons against *ab initio* references show that accurate energies and forces alone do not ensure stable NPT dynamics; correct energy–volume responses and pressure predictions are also essential. We identify two practical remedies: pressure-targeted fine-tuning and a revised NVT-quench/NPT-equilibration protocol that avoids unphysical volume expansion without additional ab initio training data. Both recover $IrO_2$ densities and local structures consistent with AIMD. Across 30 chemically diverse materials, the volume-expansion failure proves general, and the revised protocol substantially improves density predictions, reducing the AIMD-referenced MAE from 2.46 to 0.35 g/cm$^3$. This work establishes practical validation criteria and simulation strategies for robust uMLIP-driven amorphous structure generation.

## Introduction

Amorphous materials underpin a broad range of technologies, from catalysis and energy storage to optoelectronics[1,2]. Unlike crystalline phases, amorphous materials lack long-range periodicity and exhibit diverse local coordination environments, structural heterogeneity, and complex potential-energy landscapes[3]. These features can give rise to unusual and highly tunable properties, but they also make predictive atomistic modeling substantially more challenging than for their crystalline counterparts[4]. In practice, representative amorphous structures are commonly generated using melt–quench molecular dynamics (MD), which requires extensive sampling over large system sizes and long timescales. Although *ab initio* molecular dynamics (AIMD) provides a reliable first-principles route, its computational cost remains prohibitive for generating statistically meaningful amorphous models. This sampling bottleneck continues to limit the reliable modeling and rational design of amorphous materials[3].

Machine-learning interatomic potentials (MLIPs) offer a promising route to overcome this limitation by combining near-ab initio accuracy with orders-of-magnitude lower computational cost[5,6]. System-specific MLIPs have already enabled accurate large-scale simulations of amorphous systems such as carbon[7,8] and silicon[9,10]. More recently, universal MLIPs (uMLIPs), trained on large and chemically diverse datasets[11–13], have emerged as general-purpose atomistic models with broad transferability across materials classes, bonding environments, and thermodynamic conditions[14]. Their strong performance in applications ranging from elastic-property prediction[15] and defect energetics[16] to phonon calculations[17], phase-diagram exploration[18], and metal–organic framework modeling[19] has underscored their potential for accelerating atomistic simulations. However, most existing benchmarks focus on crystalline or near-equilibrium configurations, and whether uMLIPs are reliable under the strongly non-equilibrium conditions encountered during melt–quench simulations remains largely unexplored.

This raises a practical question: how can the unphysical expansion observed in uMLIP-driven NPT-MD simulations be corrected? This is a demanding task because melt–quench trajectories sample high-temperature, high-energy configurations that are poorly represented in current uMLIP training datasets. Here, we address this question using amorphous $IrO_2$ (am-$IrO_2$), a benchmark electrocatalyst for the oxygen evolution reaction in acidic water electrolysis[20,21], as a diagnostic test case. We evaluate eight leading uMLIPs (GRACE, ORB, SevenNet, MACE, MatterSim, UMA, UPET, and NequIP) under melt–quench conditions, diagnose the physical origins of their behavior, and develop practical strategies to restore realistic amorphous structure generation. We find that all examined models fail under the conventional melt–quench protocol, producing unphysically expanded structures with densities far below the AIMD reference. By comparing energies, forces, and virial pressures against *ab initio* reference calculations, we show that accurate energy–volume relations and pressure predictions across non-equilibrium states are prerequisites for reliable amorphous structure generation. We further propose and validate two remedies: targeted fine-tuning of uMLIPs with explicit emphasis on virial-pressure fidelity, and a revised simulation protocol that mitigates catastrophic expansion when fine-tuning is not feasible. More broadly, our results establish diagnostic criteria and actionable strategies for robust uMLIP-driven modeling of amorphous materials.

## Results and Discussion

Failure of uMLIPs under conventional melt–quench simulations

The melt–quench method is widely used for generating amorphous structures[22,23]. In this approach (Fig. 1a), melting is performed in the NVT ensemble, followed by quenching and equilibration in the NPT ensemble to allow volume evolution during cooling. We calculated the density of am-$IrO_2$ using eight uMLIPs (Table S1) and compared the results with those from AIMD. As shown in Fig. 1b, AIMD yields a density of 10.04 g/cm$^3$, close to the previously reported theoretical value[24] of 10.70 g/cm$^3$. This small difference can be attributed to the choice of exchange-correlation functional: the previous study used optB86b, whereas we used PBE, which tends to overestimate lattice parameters and volume, thereby giving a slightly lower density. In contrast, all eight uMLIPs (GRACE, ORB, SevenNet, MACE, MatterSim, UMA, UPET, and NequIP) predict unphysically low densities of 1–4 g/cm$^3$. Reducing the quench rate does not correct this behavior; instead, slower quenching leads to even lower predicted densities of 0.2–1.1 g/cm$^3$ across all eight uMLIPs (Fig. S1). These results indicate that current uMLIPs fail to reproduce the formation of a physically realistic am-$IrO_2$ phase during melt–quench simulations.

Importantly, this failure is not unique to $IrO_2$. To assess its generality, we selected 30 representative materials, including binary oxides, nitrides, sulfides, halides and ternary oxides, and performed melt–quench MD simulations. Of these, 19 produced densities substantially below the corresponding reference densities (Table S2), indicating that unphysical expansion is a generic problem in uMLIP-driven amorphous structure generation.

Fig. 1c and 1d show the density and structural evolution during MD quenching. Using GRACE as a representative example, we observe that the density drops sharply to below 1.00 g/cm$^3$ within the first 10 ps of quenching, indicating strong volumetric expansion. The structural snapshots confirm this picture by showing rapid expansion of the system: by 1.0 ps the structure already appears gas-like, and by 10 ps it has become a highly dispersed configuration with no condensed-phase continuity. Together, these observations indicate that uMLIP simulations do not form a stable amorphous network during cooling, but instead evolve toward an unphysical low-density state.

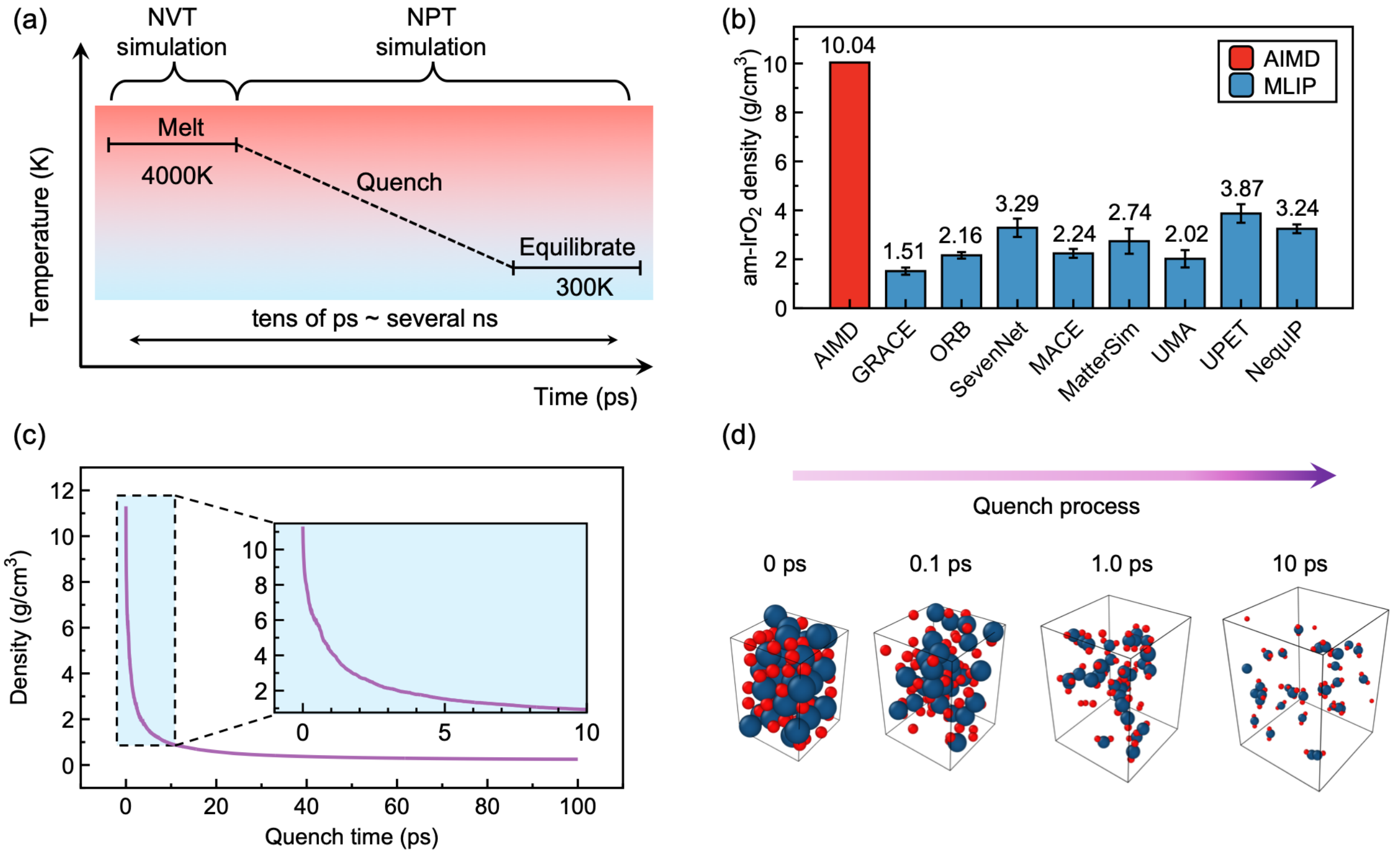

**Fig. 1 | Failure of uMLIPs to reproduce amorphous $IrO_2$ density under the conventional melt–quench protocol**. (a) Schematic illustration of amorphous structures generated by a typical melt–quench MD simulation. (b) Density of am-$IrO_2$ obtained from melt–quench MD simulations using AIMD and uMLIPs at a quench rate of 580 K/ps. Each error bar represents the average density from five independent simulations, and the black error bars indicate the corresponding standard deviation. (c) Temporal evolution of the am-$IrO_2$ density during quenching, obtained from GRACE MD simulations with an extended quench time of 100 ps, corresponding to a quench rate of 37 K/ps. (d) Selected structural snapshots illustrating the structural evolution during quenching. Blue and red spheres represent iridium and oxygen atoms, respectively.

## Energy–volume and pressure errors behind unphysical expansion

To understand the origin of this anomalously large volume expansion, we extracted configurations every 0.01 ps from the first 1.0 ps of the quench trajectory and benchmarked their energies, forces, and pressures against density functional theory (DFT) calculations (Fig. 2). We also computed these quantities for the same configurations using our previously developed moment tensor potential (MTP)[25], a system-specific MLIP reference that reproduces the AIMD behavior of am-$IrO_2$ (see Methods). In contrast to the uMLIPs, the MTP predicts an amorphous density of 10.20

g/cm$^3$, close to the AIMD value of 10.04 g/cm$^3$, and shows little sensitivity to the quench rate (Fig. S2). Thus, comparison with the MTP helps identify which model properties distinguish successful from failed NPT quenches.

GRACE tracks the DFT reference energies more closely than the MTP throughout the quench (Fig. 2a). The MTP–DFT energy difference increases with quench time, reaching approximately 0.60 eV/atom at 1.0 ps (Fig. 2b). This indicates that the MTP systematically overestimates the energy as the system expands into the low-density regime. By contrast, GRACE remains much closer to DFT in absolute magnitude, but its error drifts from near zero to approximately −0.08 eV/atom over the same interval (Fig. 2c), indicating that GRACE increasingly underestimates the energy as the density decreases. The two MLIPs therefore exhibit opposite density-dependent energy errors under expansion: the MTP increases the energetic cost of volume growth, whereas GRACE reduces it. This difference, rather than the absolute magnitude of the energy error, is central to their distinct quench behavior.

The mean-force discrepancies are smaller than the energy discrepancies. Both GRACE and MTP broadly reproduce the DFT force evolution during the first 1.0 ps (Fig. 2d). The MTP force deviations fluctuate around the DFT values, with increased scatter at later times, but remain within ~0.50 eV/Å (Fig. 2e), whereas GRACE exhibits smaller deviations, within ~0.15 eV/Å (Fig. 2f). Thus, in terms of instantaneous mean-force accuracy, GRACE outperforms the am-$IrO_2$-specific MTP.

The pressures also track the DFT trend qualitatively for both MLIPs but differ in magnitude (Fig. 2g). The MTP exhibits a larger pressure deviation that grows with quench time, reaching ~$6 \times 10^4$ bar (Fig. 2h), whereas GRACE gives smaller deviations, generally within ~$1.5 \times 10^4$ bar (Fig. 2i). Thus, for the early-stage quench configurations, GRACE shows smaller instantaneous errors than the MTP in energy, mean-force, and pressure (Fig. S3). Nevertheless, GRACE drives a pronounced density decrease during NPT quenching, whereas MTP does not.

These results reveal an apparent paradox: MTP, the less accurate model on these instantaneous metrics, produces the physically correct density, whereas GRACE, the more accurate model, drives the unphysical expansion. The resolution lies in the sign of the energy–volume (E–V) error rather than its magnitude. The MTP's positive error penalizes volume expansion and preserves the correct density, whereas GRACE's negative error promotes further expansion and destabilizes the system. An MLIP can therefore be more accurate than a domain-specific reference on every instantaneous quantity yet still fail catastrophically under NPT dynamics if the sign of its E–V response is wrong. Long-time stability in NPT quench simulations is governed not by pointwise agreement with DFT on isolated snapshots, but by the fidelity of the density-dependent energy response under expansion.

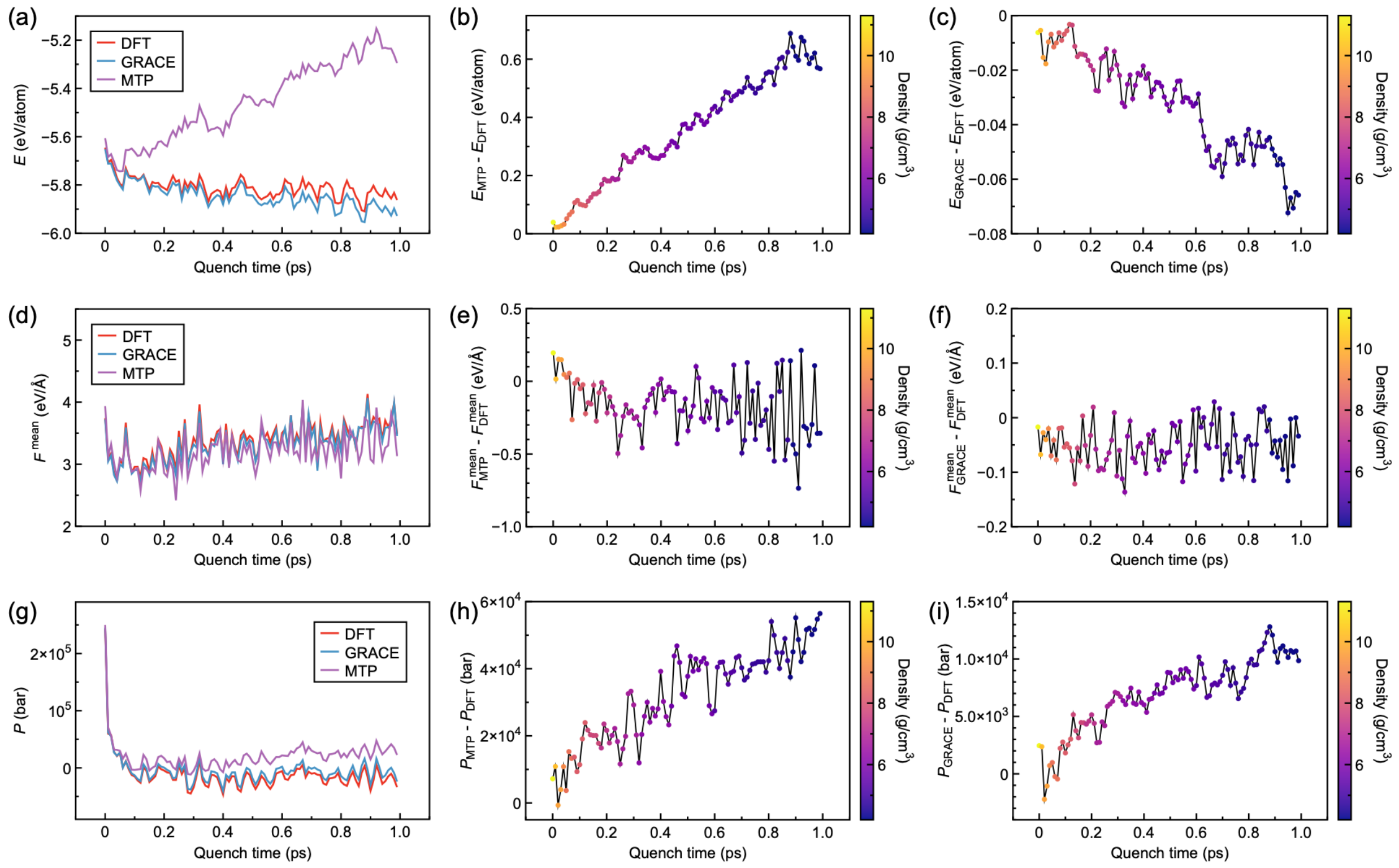

**Fig. 2 | Per-snapshot benchmarking of GRACE and MTP during early-stage NPT quenching**. Temporal evolution of energy ($E$), mean force ($F^{\mathrm{mean}}$), and virial pressure ($P$) during the first 1.0 ps of quenching is shown in panels (a), (d) and (g), respectively. Panels (b,c), (e,f), and (h,i) show the corresponding deviations of MTP and GRACE from DFT for energy, mean force, and virial pressure, respectively.

To further examine the E–V relationship, we calculated the energies of approximately 2,400 configurations spanning a range of densities, or equivalently volumes, sampled during NPT-MD quenching using eight uMLIPs. These results were benchmarked against DFT and MTP calculations, as shown in Fig. 3. For the DFT-calculated E–V curve, the system energy initially decreases, reaches a minimum at approximately 9.0–10.0 g/cm$^3$, and then increases as the system continues to expand. MTP exhibits a similar trend but significantly overestimates the energy in the low-density region relative to DFT. In contrast, GRACE, ORB, SevenNet, MACE, MatterSim, and UMA predict a monotonic decrease in system energy with decreasing density (Fig. 3c–h). Consequently, during NPT simulations, these uMLIPs incorrectly predict abnormally low-density states to be energetically favorable. Interestingly, UPET and NequIP exhibit E–V relationships that closely match the DFT results (Fig. 3i–j), reproducing both the characteristic energy minimum and a reasonable increase in energy upon expansion. The absolute energy errors of UPET and NequIP are significantly smaller than those of MTP across the entire density range. However, NPT-MD simulations using these two uMLIPs still yield unphysically low densities of 3.87 g/cm$^3$ for UPET and 3.24 g/cm$^3$ for NequIP in am-$IrO_2$ (Fig. 1b).

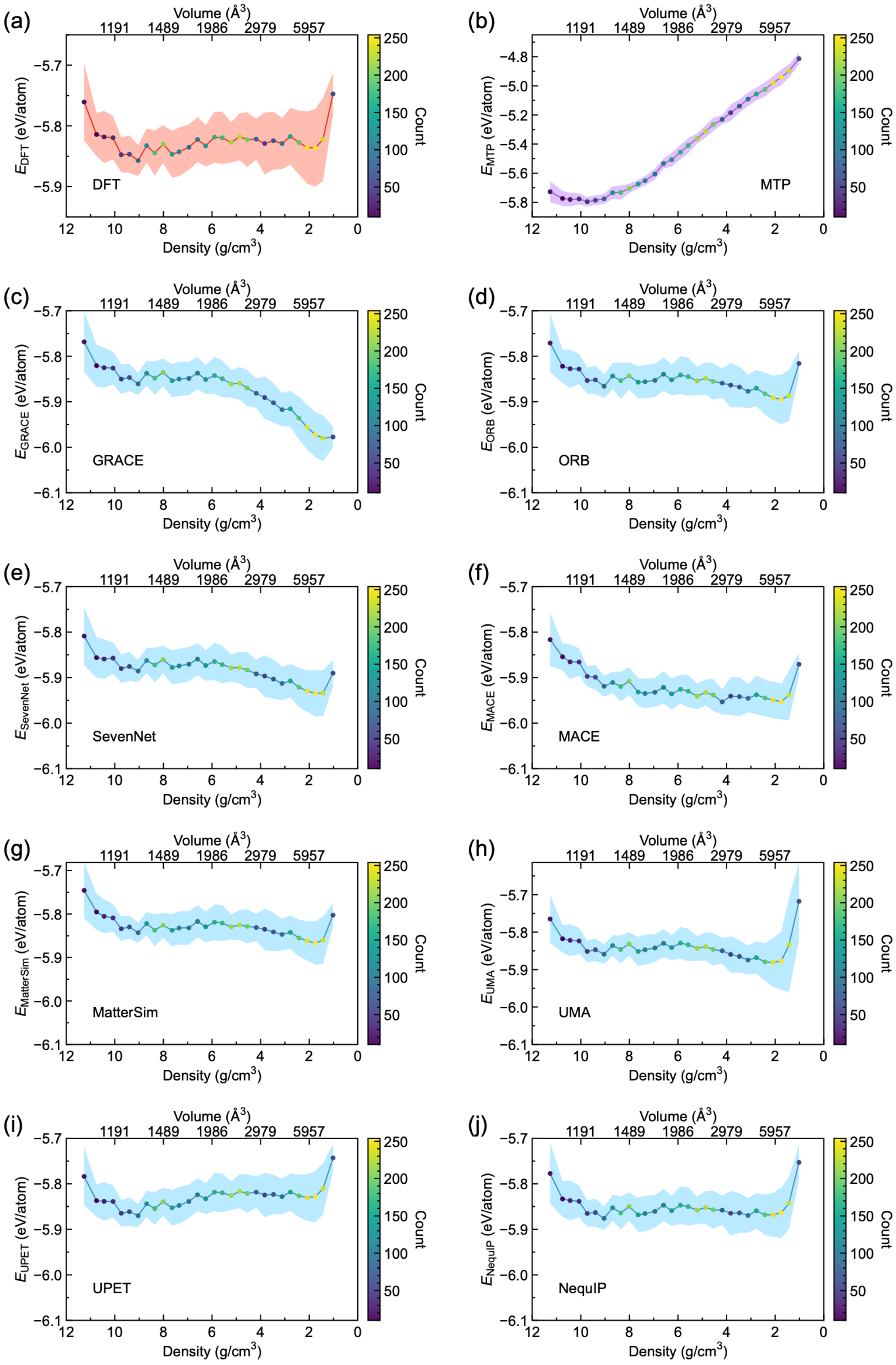

**Fig. 3 | Energy–volume responses of amorphous $IrO_2$ from DFT, MTP, and uMLIPs**. Energy of am-$IrO_2$ as a function of density (bottom *x*-axis) and volume (top *x*-axis) from (a) DFT, (b) MTP, and (c–j) eight uMLIPs. Shaded regions indicate the standard deviation, and the color map denotes the sampling count at each density.

To investigate the origin of this discrepancy, we benchmarked the energies, forces, and pressures predicted by UPET and NequIP against AIMD reference values. UPET accurately reproduces the AIMD energies, with a mean absolute error (MAE) of 44.6 meV/atom and an $R^2$ of 0.963 (Fig. 4a). Its force predictions also agree closely with AIMD, yielding an MAE of 0.029 eV/Å and an $R^2$ of 0.998 (Fig. 4b). The pressure, however, is poorly reproduced: although the UPET-predicted pressures broadly follow the AIMD trend, their magnitudes are substantially overestimated, resulting in a large MAE of 79,414.2 bar and a negative $R^2$ of −9.574 (Fig. 4c). NequIP exhibits the same behavior (Fig. S4), reproducing energies and forces accurately but systematically overestimating pressure. This shared deficiency in pressure prediction likely explains the unphysical expansion observed in NPT-MD simulations with both models.

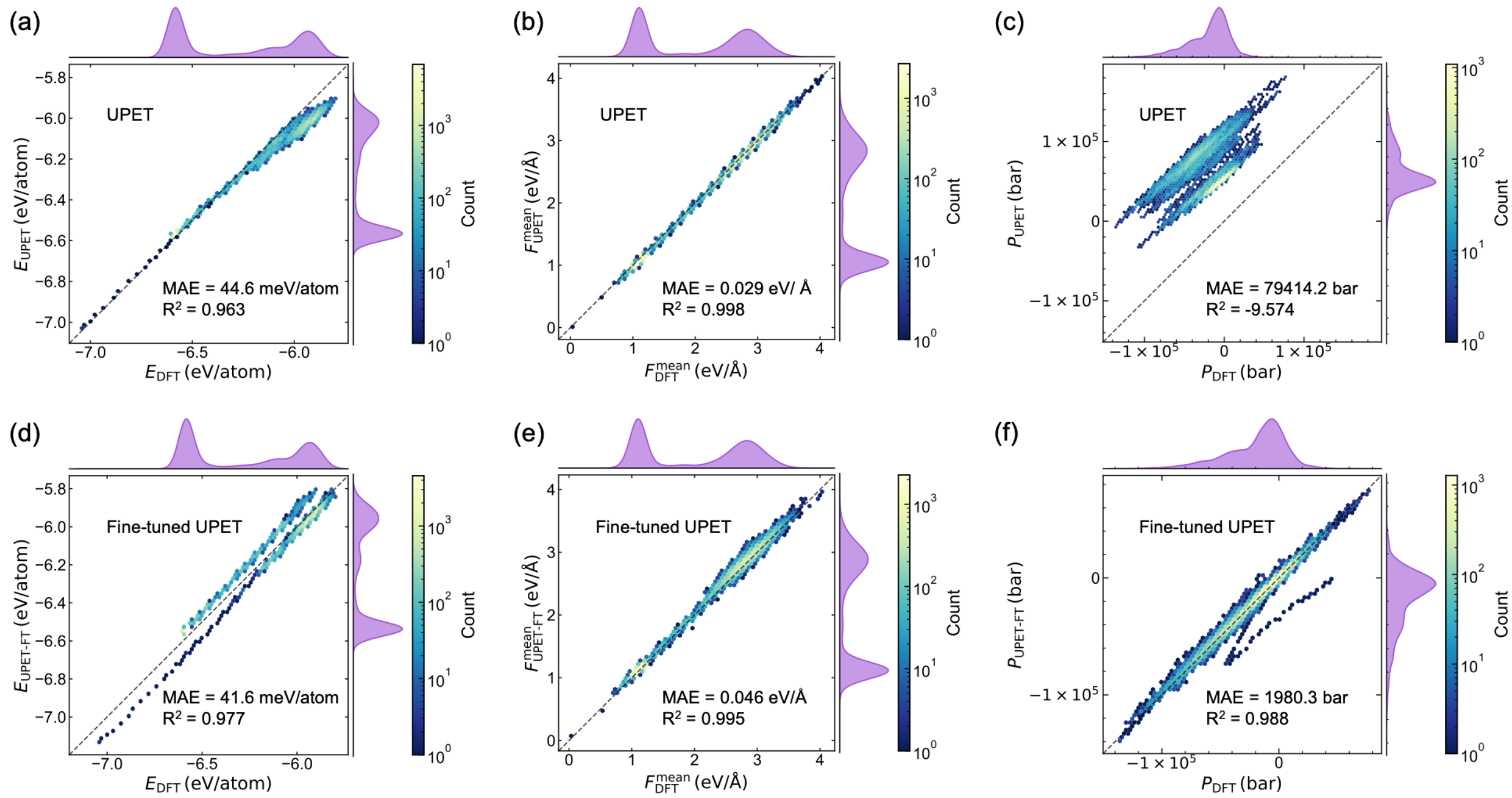

**Fig. 4 | Pressure-targeted fine-tuning of UPET for improved pressure prediction**. Parity plots of (a–c) UPET and (d–f) fine-tuned UPET (UPET-FT) predictions against AIMD reference values for energy [(a), (d)], mean force [(b), (e)], and pressure [(c), (f)], using configurations sampled during AIMD. The color map denotes the sampling count in each bin.

## Pressure-targeted fine-tuning of uMLIPs

This raises a critical question: how can the unphysical expansion observed in uMLIP-driven NPT-MD simulations be corrected? The most straightforward approach is to fine-tune the foundation potential models. Taking UPET as an example, we first tested whether pressure-targeted fine-tuning could improve NPT quenching performance. During fine-tuning, the stress component was strongly weighted ($w_S$ = 10.0) to emphasize pressure correction, whereas the already accurate energy and force terms were downweighted ($w_E$ = 0.1, $w_F$ = 0.1) (see Methods). The results for the fine-tuned model, UPET-FT, are shown in Fig. 4d–f. UPET-FT preserves excellent agreement with AIMD for energies, with an MAE of 41.6 meV/atom and an $R^2$ of 0.977, and remains accurate for forces, with an MAE of 0.046 eV/Å and an $R^2$ of 0.995. Crucially, its pressure predictions are substantially improved, yielding an MAE of 1,980.3 bar and an $R^2$ of 0.988 relative to AIMD.

We next evaluated UPET-FT by computing the density of am-$IrO_2$ and comparing it with the AIMD reference and the original UPET prediction (Fig. 5a). UPET-FT yields 10.66 g/cm$^3$, in good agreement with the AIMD value of 10.04 g/cm$^3$ and a substantial improvement over the 3.87 g/cm$^3$ predicted by the original UPET. This improved accuracy also extends to the local atomic structure. The UPET-FT radial distribution functions (RDFs, $g(r)$) reproduce the characteristic Ir–O, O–O, and Ir–Ir peak positions at 1.96, 2.76, and 3.48 Å, respectively, in excellent agreement with the AIMD reference (Fig. 5b and 5c). These results confirm that pressure-targeted fine-tuning enables the uMLIP to accurately reproduce both the density and the local structure of am-$IrO_2$.

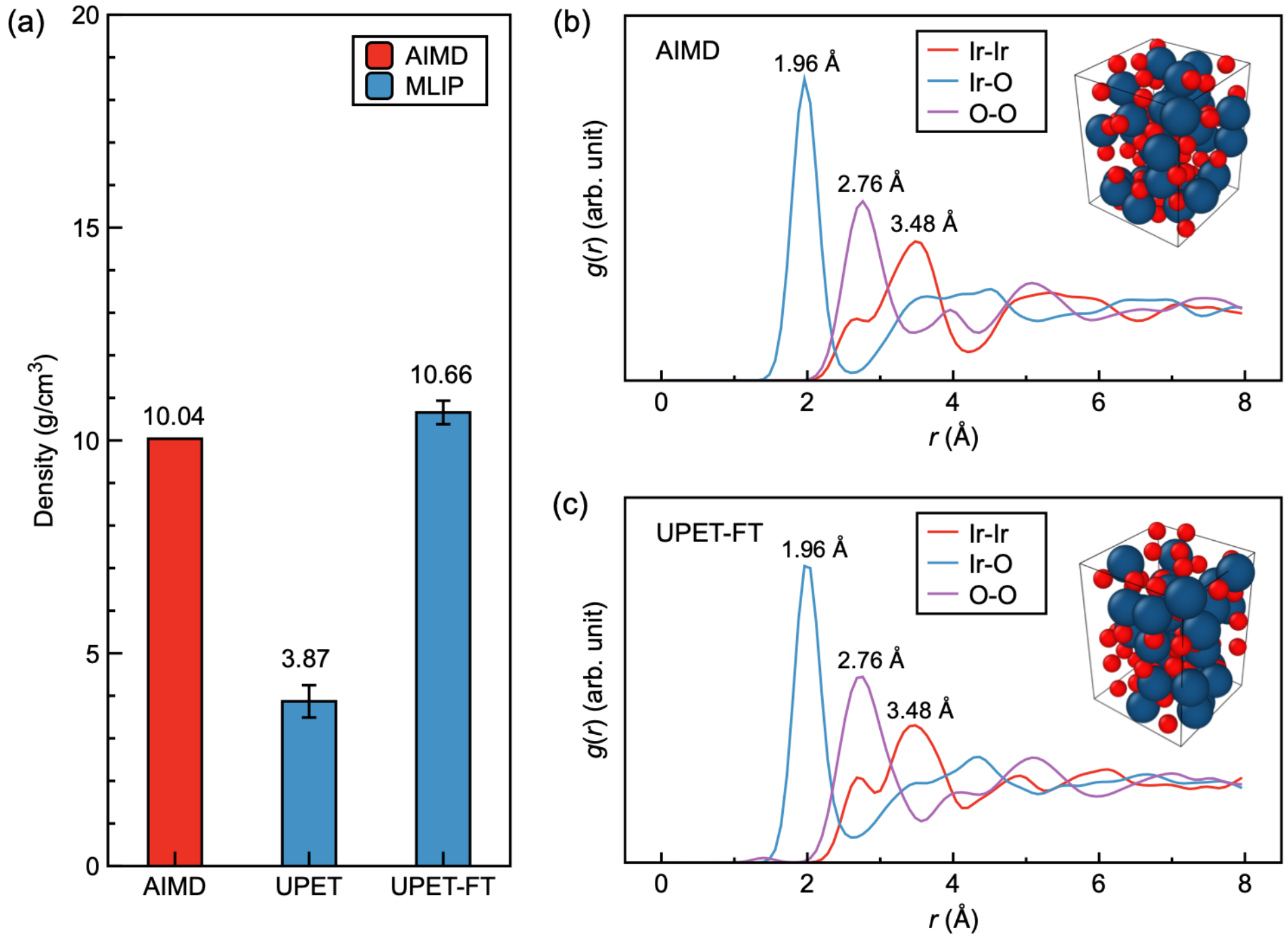

**Fig. 5 | Density and local-structure recovery by pressure-targeted UPET fine-tuning**. (a) Densities of am-$IrO_2$ obtained from AIMD, the original UPET, and fine-tuned UPET (UPET-FT). Each error bar represents the average density from five independent simulations, and the black error bars indicate the corresponding standard deviation. RDFs of am-$IrO_2$ from (b) AIMD and (c) UPET-FT, with insets showing the atomic structures.

Revised NVT-quench/NPT-equilibration protocol

Although fine-tuning is an effective strategy for improving large foundation models, it requires sufficient complementary training data and thus additional computational effort. We therefore developed a revised melt–quench protocol that avoids the need for additional fine-tuning data. The original UPET model shows a pronounced temperature dependence in its pressure error, with substantially smaller errors at low temperature than at high temperature (Fig. 6a). Notably, the low-temperature pressure error is comparable to the pressure MAE achieved by the fine-tuned UPET model (Fig. 4f), and the same trend is observed for the other tested uMLIPs (Fig. S5). This motivates a modified quenching protocol in which NVT simulations are used for the high-temperature quench, while NPT simulations are applied only during low-temperature equilibration (Fig. 6b). Because NVT quenching fixes the cell volume, high-temperature energy–volume errors no longer drive volume evolution during the quench. Using this protocol, all tested uMLIPs produce densities of ~10.00 g/cm$^3$, in good agreement with the AIMD reference value (Fig. 6c). The resulting amorphous structures also show good agreement with the AIMD reference, as evidenced by the RDFs (Fig. 6d). The Ir–O, O–O, and Ir–Ir RDF peaks predicted by GRACE occur at 1.96, 2.76, and 3.56 Å, respectively, closely matching the corresponding AIMD peak positions (Fig. 5b). The RDFs from other uMLIP models are provided in Fig. S6. These results indicate that the revised NVT-quench protocol provides a practical route to amorphous models that accurately reproduce both the density and local structure of am-$IrO_2$, without requiring model fine-tuning.

Beyond $IrO_2$, we extended the revised protocol to all 30 benchmark materials and compared the predicted amorphous densities with reference values taken from AIMD, experiment, or the corresponding crystalline phases (Table S2). As shown in Fig. 6e, the densities predicted by GRACE using the revised NVT-quench protocol agree closely with the reference values, which span 1–8 g/cm$^3$, giving an MAE of 0.50 g/cm$^3$. When the comparison is restricted to AIMD reference values, the MAE decreases to 0.35 g/cm$^3$, providing a more meaningful metric because uMLIP-predicted properties are most appropriately evaluated against AIMD-level reference data. In contrast, the conventional NPT-quench protocol yields a substantially larger MAE of 2.95 g/cm$^3$, or 2.46 g/cm$^3$ when restricted to AIMD references. This degradation arises primarily because most materials undergo unphysical volume expansion during NPT quenching, producing gas-like densities with absolute relative errors exceeding 90% (Fig. 6f). This broad benchmark validation further demonstrates the robustness of the revised protocol for uMLIP-driven amorphous structure modeling.

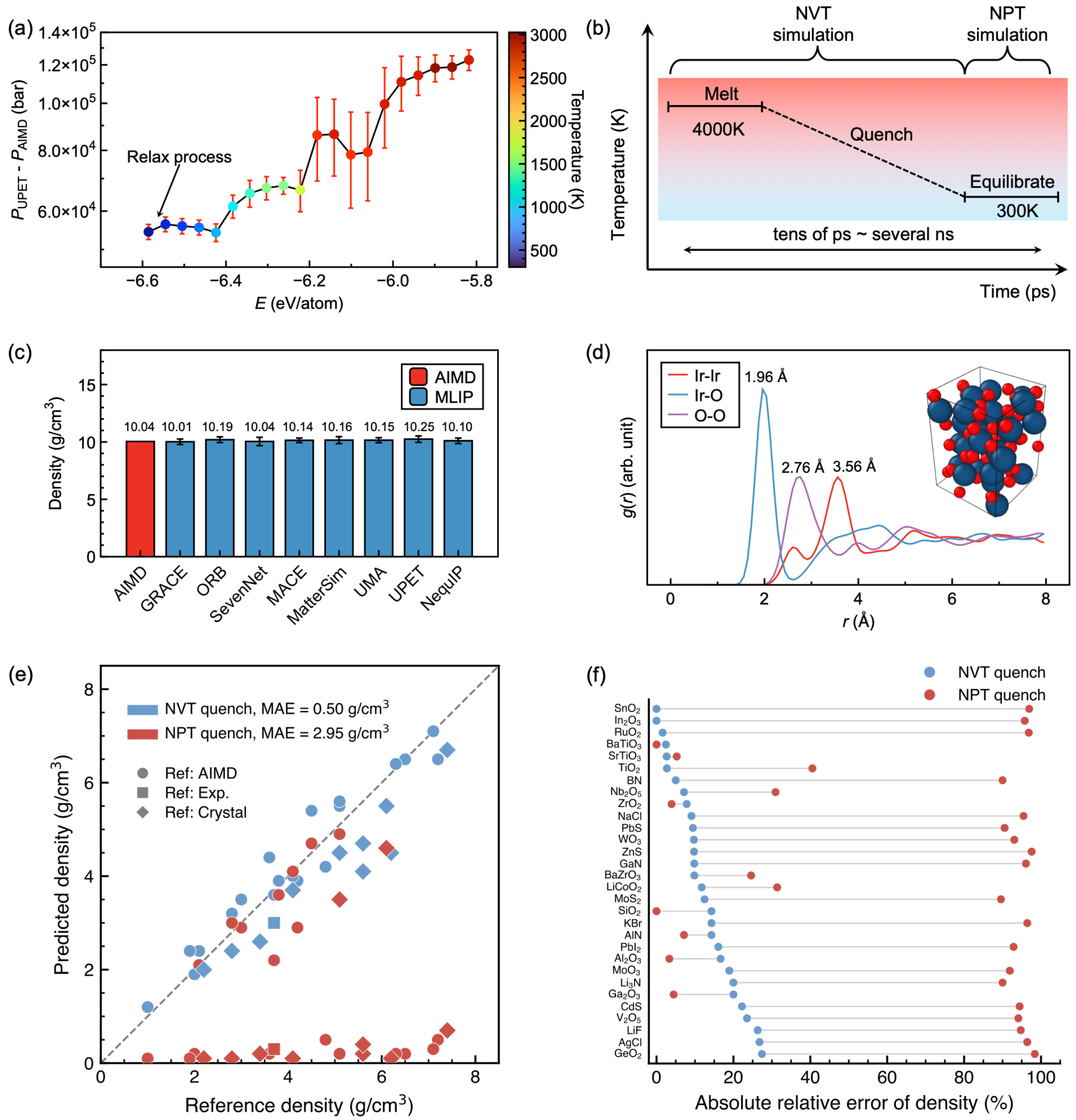

**Fig. 6 | Revised melt–quench protocol for uMLIP-driven amorphous structure generation.** (a) Pressure difference between the original UPET and AIMD for $IrO_2$ as a function of energy, with points colored by temperature. Data points and error bars represent the mean and standard deviation within each energy bin, respectively. (b) Schematic illustration of the revised melt–quench protocol. (c) Densities of am-$IrO_2$ obtained from AIMD and the original eight uMLIPs using the revised protocol. Each error bar represents the average density from five independent simulations, and the black error bars indicate the corresponding standard deviation. (d) RDFs of

the am-$IrO_2$ structure generated by GRACE using the revised protocol, with an inset showing the atomic structure. (e) Amorphous densities of the 30 benchmark materials predicted by GRACE using the revised NVT-quench (blue) and NPT-quench (red) protocols, plotted against their reference densities. Marker shapes denote the source of the reference density: circles for AIMD, squares for experiment, and diamonds for the corresponding crystalline phase. (f) Absolute relative errors of the predicted densities for each material under the revised NVT-quench (blue) and NPT-quench (red) protocols.

## Conclusion

In summary, we have systematically evaluated eight leading uMLIPs for generating amorphous $IrO_2$ through melt–quench MD simulations. Although several models show good agreement with DFT for energies and forces, all tested uMLIPs fail under the conventional melt–quench protocol, producing unphysically expanded structures with densities far below the AIMD reference. This failure extends beyond $IrO_2$, generalizing across a 30-material benchmark and indicating that unphysical expansion is a broad challenge in uMLIP-driven amorphous structure generation. Our analysis identifies two key origins of this failure: erroneous E–V behavior, which incorrectly favors low-density configurations, and inaccurate pressure predictions, which drive incorrect volume evolution even when the E–V relationship is reasonable. We further demonstrate two practical solutions. Pressure-targeted fine-tuning of UPET substantially reduces the pressure error and restores both the density and local structure of am-$IrO_2$ in agreement with AIMD. Alternatively, a revised protocol using NVT quenching followed by low-temperature NPT equilibration produces realistic amorphous structures for all tested uMLIPs without additional fine-tuning data. The 30-material benchmark further confirms the broader applicability of this strategy, showing that the revised protocol suppresses unphysical volume expansion and reduces the mean absolute density error from 2.46 to 0.35 g/cm$^3$ across diverse chemistries. Overall, our results show that reliable uMLIP-driven amorphous structure modeling requires validation beyond conventional energy and force benchmarks, with E–V behavior and pressure accuracy serving as essential criteria for assessing thermodynamic stability under melt–quench conditions.

# Methods

## Molecular Dynamics Simulations

**uMLIP-driven molecular dynamics**: Melt–quench molecular dynamics (MD) simulations driven by universal machine-learning interatomic potentials (uMLIPs) were performed using the Atomic Simulation Environment (ASE)[26]. Temperature was controlled with a Langevin thermostat in the canonical (NVT) ensemble, whereas pressure control in the isothermal–isobaric (NPT) ensemble was achieved using the Berendsen-style barostat implemented in ASE. Unless otherwise stated, a time step of 1 fs was used for all MD simulations. Energies and forces were evaluated using the corresponding pretrained uMLIP models interfaced with ASE, including GRACE[27], ORB[28], SevenNet[29], MACE[30], MatterSim[31], UMA[32], UPET[33], and NequIP[34].

**MTP-driven molecular dynamics**: Moment tensor potential (MTP) simulations were performed using LAMMPS[35]. The MTP was developed with the MLIP-3[36] package through an active learning workflow. Initial training configurations were sampled from AIMD trajectories of both crystalline rutile $IrO_2$ (120-atoms) and disordered $IrO_2$ (96-atoms) phases over a temperature range of 300–2500 K and applied strains from -0.05 to +0.05. From these trajectories, 1,800 representative snapshots were selected for high-fidelity DFT static calculations of energies, forces, and stresses. The MTP was then refined using a learning-on-the-fly (LOTF) protocol with a 5.0 Å cutoff and $lev_{max}$ = 16. The relative weights assigned to energy, force and stress were 1:0.001:0.01. Active learning was performed systematically in both NVT and NPT ensembles, covering melting at 2500 K, controlled cooling to 300 K, and final equilibration. Further details of the MTP training and validation procedure are provided in Ref. 25. Because the MTP workflow is implemented through LAMMPS rather than an ASE calculator, all MTP-driven melt–quench simulations were performed separately in LAMMPS.

**Melt–quench protocol**: The melt–quench protocol used for the uMLIP- and MTP-driven simulations is summarized in Fig. 1a and 6b. For $IrO_2$, the simulation was initialized from the crystalline structure. For the other benchmark materials, initial configurations were generated as random atomic packings using Packmol[37], with the packing density set to 1.05 times that of the ground-state crystalline phase obtained from Materials Project[38]. In each simulation, the initial model was first melted at 4000 K for 10 ps, then quenched to 300 K at the prescribed cooling rate, and finally equilibrated at 300 K for 10 ps. The cooling rates used for $IrO_2$ in the different simulations are provided in the corresponding figure captions, whereas a cooling rate of 37 K/ps was used for all 30 benchmark materials.

## Ab initio Molecular Dynamics and DFT Reference Calculations

**AIMD simulations**: Ab initio molecular dynamics (AIMD) simulations were performed using the Vienna *Ab initio* Simulation Package (VASP)[39,40]. The Perdew-Burke-Ernzerhof (PBE) functional[41] was used to describe exchange–correlation interactions. A plane-wave energy cutoff of 400 eV was applied. The Brillouin zone was sampled at the $\Gamma$ point. The electronic energy convergence criterion was set to $10^{-5}$ eV. Reference amorphous structures were generated following the melt–quench protocol of Lee et al.[24]. Initial supercells containing 96 atoms were simulated in the NPT ensemble using Parrinello–Rahman dynamics combined with a Langevin thermostat. The systems were melted at 3000 K for 10 ps, linearly quenched from 3000 to 100 K over 5 ps, and subsequently equilibrated at 300 K for 10 ps. A 1 fs time step was used throughout the AIMD simulations.

**DFT reference calculations**: Static DFT calculations were performed for configurations sampled from the uMLIP-MD trajectories to provide reference energies, forces, and stresses for benchmarking. These calculations used VASP with the PBE functional and a plane-wave energy cutoff of 520 eV. To maintain a comparable reciprocal-space sampling density across configurations with different cell volumes, the Monkhorst–Pack $k_1 \times k_2 \times k_3$ meshes were selected such that $k_i|a_i| \geq 20$ Å for each lattice direction, where $|a_i|$ is the length of lattice vector $a_i$ and $k_i$ is the number of $k$ points along that direction. The electronic convergence criterion was set to $10^{-5}$ eV.

## MLIP Benchmarking and Fine-Tuning

**Benchmarking metrics**: The uMLIPs and MTP were benchmarked against DFT and AIMD reference data using three configuration-level quantities relevant to melt–quench simulations: the energy per atom ($E$), the mean atomic force magnitude ($F^{\mathrm{mean}}$), and the virial pressure ($P$) derived from the stress tensor. These quantities were defined as:

$$E = E_{\mathrm{tot}}/N \quad (1)$$

$$F^{\mathrm{mean}} = \frac{1}{N}\sum_{i=1}^{N} |F_i| \quad (2)$$

$$P = -\frac{1}{3}\mathrm{Tr}(\sigma) \quad (3)$$

where $E_{\mathrm{tot}}$ is the total energy of a structure, $N$ is the number of atoms, $F_{\mathrm{i}}$ is the force vector acting on atom $i$, and $\sigma$ is the stress tensor. The sign convention for pressure follows Equation (3).

**UPET fine-tuning**: UPET was fine-tuned from the pretrained checkpoint *pet-oam-xl-v1.0.0.ckpt* using configurations extracted from the AIMD trajectory. The dataset was randomly divided into training, validation, and test subsets at a ratio of 0.8:0.1:0.1. Fine-tuning was performed for 100 epochs with a learning rate of $1 \times 10^{-5}$. The optimization objective combined energy, force, and stress losses according to

$$L = 0.1L_E + 0.1L_F + 10.0L_\sigma \quad (4)$$

The stress term was assigned a larger weight to improve the prediction of pressure and volume evolution during melt–quench simulations. The energy and force terms were assigned smaller weights to preserve the accuracy of the pretrained model for these quantities during fine-tuning.

**Structural analysis and visualization**: Radial distribution functions (RDFs) were calculated using OVITO[42]. Atomic configurations were rendered in OVITO using the element-specific color schemes described in the corresponding Figure captions.

## Data Availability

All the data supporting this study are included in the article and its supplementary information.

## Acknowledgements

This work was supported by the City University of Hong Kong Start-up Grant (No. 9020004). Some calculations were performed on the CityU Burgundy computational facilities, managed and provided by the Computing Services Centre at the City University of Hong Kong.

## Author Contributions

Z.W. conceived the project and supervised the research. S.L. and Y.A. performed the calculations, analyzed the data, and wrote the original draft. X.G. contributed to the data analysis. W.Y. performed the UMA calculations. Z.W. reviewed and edited the manuscript. All authors discussed the results and commented on the manuscript.

## Competing Interests

The authors declare no competing interests.

## Additional information

Additional supporting information can be found online in the Supporting Information section.

# Supplementary Information

# Melt–Quench Failures and Practical Solutions for Universal Machine-Learning Interatomic Potentials in Amorphous Structure Generation

Shuwei Li[1†], Yuqi An[1†], Xingyu Guo[2], Wenqiang Yang[3*], Zhenbin Wang[1,4*]

[1]Department of Materials Science and Engineering, City University of Hong Kong, Hong Kong SAR, 999077, China

[2]Department of Data Science, City University of Hong Kong, Hong Kong SAR, 999077, China

[3]Department of Chemical Engineering, University of South Carolina, Columbia, South Carolina, USA

[4]School of Energy and Environment, City University of Hong Kong, Hong Kong SAR, 999077, China

[†]These authors contributed equally to this work.

*Corresponding author, E-mail: wenqiang@email.sc.edu; zwan22@cityu.edu.hk

**Supplementary Table 1 | Universal machine-learning interatomic potentials (uMLIPs) used in this study**. Model variants and training-data sources for the uMLIPs benchmarked for amorphous $IrO_2$ generation.

| uMLIPs | Model | Training data source |
|---|---|---|
| GRACE | GRACE-2L-OMAT[1] | OMat24[2] |
| ORB | ORB-v3[3] | MPtrj[4]+Alex[5]+OMat24[2] |
| SevenNet | SevenNet-Omin-i12[6] | COSMOS[6] (15 publicly available *ab initio* databases) |
| MACE | MACE-MPA-0[7] | MPtrj[4]+sAlex[2,5] |
| MatterSim | MatterSim-v1-5M[8] | MatterSim[8] |
| UMA | UMA-S[9] | OMol25[10]+OMat24[2]+OC20[11]+ODAC25[12]+OMC25[13] |
| UPET | PET-OAM-XL[14] | OMat24[2]+sAlex[2,5]+MPtrj[4] |
| NequIP | Nequip-OAM-XL[15] | OMat24[2]+sAlex[2,5]+MPtrj[4] |

**Supplementary Table 2 | Comparison of amorphous densities predicted by GRACE against reference values for 30 diverse materials.** For each material, the table lists the chemical formula, the GRACE-predicted density from the NVT-quench and NPT-quench methods, the reference density, and the source of that reference. Reference values are selected by availability in the following priority order: (1) densities from ab initio molecular dynamics at the PBE level unless otherwise noted (AIMD), (2) experimental amorphous densities obtained from the literature (Exp.), and (3) crystalline-phase densities from the Materials Project[16] (Crystal).

| Materials | NVT quench Predicted density ($g/cm^3$) | NPT quench Predicted density ($g/cm^3$) | Reference density ($g/cm^3$) | Reference source |
|---|---|---|---|---|
| $SiO_2$ | 2.4 | 2.1 | 2.1 | AIMD[17] |
| $Al_2O_3$ | 3.5 | 2.9 | 3.0 | AIMD[18] |
| $TiO_2$ | 3.6 | 2.2 | 3.7 | AIMD[19] |
| $SnO_2$ | 6.5 | 0.2 | 6.5 | AIMD[20] |
| $WO_3$ | 6.5 | 0.5 | 7.2 | AIMD*[21] |
| $ZrO_2$ | 5.5 | 4.9 | 5.1 | AIMD[17] |
| $In_2O_3$ | 7.1 | 0.3 | 7.1 | AIMD[22] |
| $Nb_2O_5$ | 3.9 | 2.9 | 4.2 | AIMD[23] |
| $Ga_2O_3$ | 5.4 | 4.7 | 4.5 | AIMD[24] |
| $RuO_2$ | 6.4 | 0.2 | 6.3 | AIMD[25] |
| $SrTiO_3$ | 3.9 | 3.6 | 3.8 | AIMD[26] |
| $BaTiO_3$ | 4.0 | 4.1 | 4.1 | AIMD[27] |
| GaN | 5.6 | 0.2 | 5.1 | AIMD[28] |
| $Li_3N$ | 1.2 | 0.1 | 1.0 | AIMD[29] |
| AlN | 3.2 | 3.0 | 2.8 | AIMD[30] |
| BN | 1.9 | 0.2 | 2.0 | AIMD[31] |
| $MoS_2$ | 4.2 | 0.5 | 4.8 | AIMD[32] |
| CdS | 4.4 | 0.2 | 3.6 | AIMD[29] |
| LiF | 2.4 | 0.1 | 1.9 | AIMD[29] |
| $MoO_3$ | 3.0 | 0.3 | 3.7 | Exp.[33] |
| $V_2O_5$ | 2.6 | 0.2 | 3.4 | Crystal |
| $GeO_2$ | 4.5 | 0.1 | 6.2 | Crystal |
| $LiCoO_2$ | 4.5 | 3.5 | 5.1 | Crystal |
| $BaZrO_3$ | 5.5 | 4.6 | 6.1 | Crystal |
| PbS | 6.7 | 0.7 | 7.4 | Crystal |
| ZnS | 3.7 | 0.1 | 4.1 | Crystal |
| NaCl | 2.0 | 0.1 | 2.2 | Crystal |
| AgCl | 4.1 | 0.2 | 5.6 | Crystal |
| KBr | 2.4 | 0.1 | 2.8 | Crystal |
| $PbI_2$ | 4.7 | 0.4 | 5.6 | Crystal |
| AIMD only (n = 19) | | | | |
| MAE (g/cm$^3$) | 0.35 | 2.46 | | |
| RMSE (g/cm$^3$) | 0.44 | 3.51 | | |
| All (n = 30) | | | | |
| MAE (g/cm$^3$) | 0.50 | 2.95 | | |
| RMSE (g/cm$^3$) | 0.64 | 3.77 | | |

***Note:** The AIMD reference density was obtained using the LDA functional.

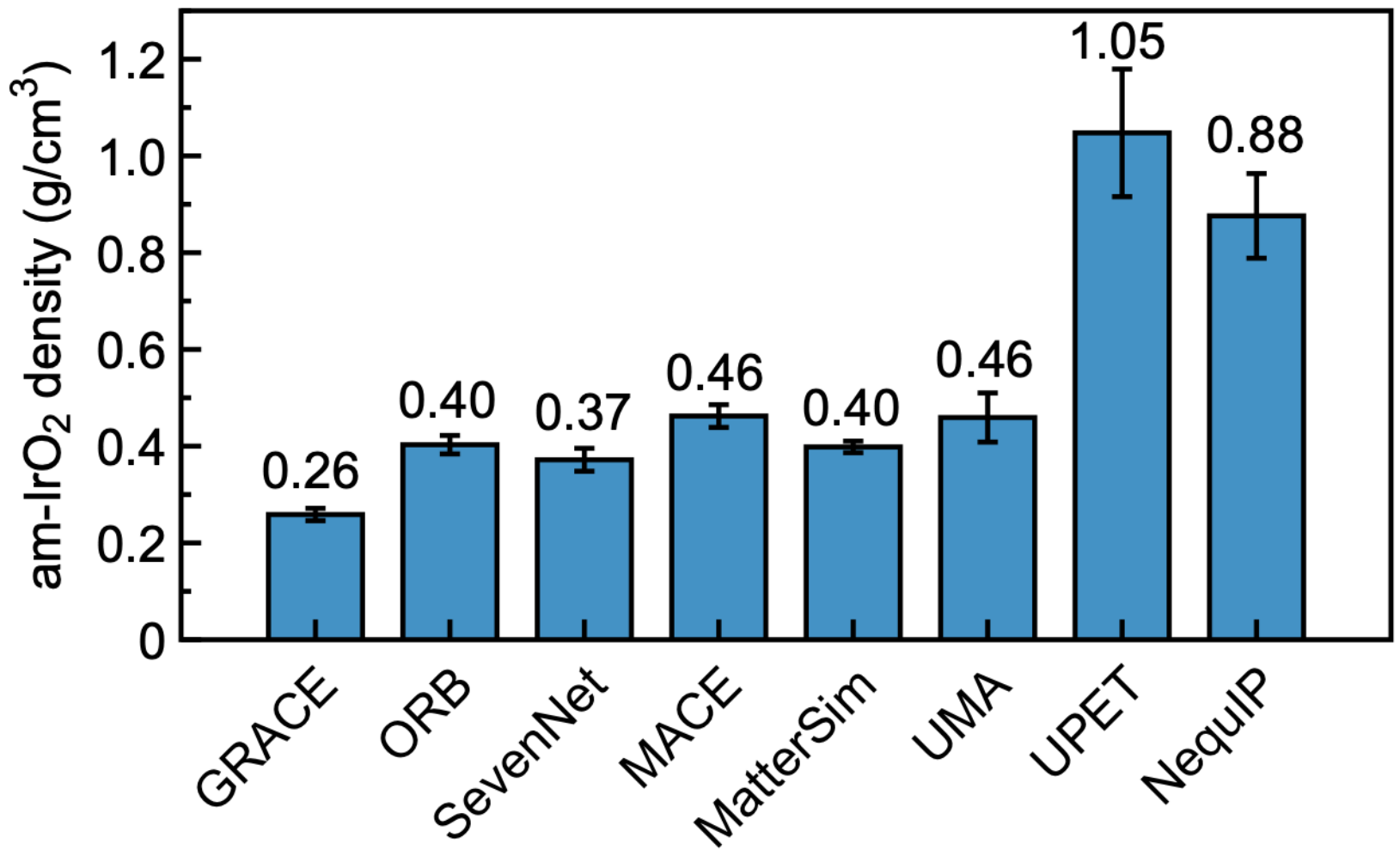

**Supplementary Fig. 1 | Density of am-$IrO_2$ generated by conventional melt-quench simulations.** Final density of am-$IrO_2$ structures generated using eight uMLIPs in conventional melt-quench MD simulations. Simulations used a quench rate of 37 K/ps and a quench time of 100 ps. Each bar represents the average density from five independent simulations, and the black error bars indicate the corresponding standard deviation.

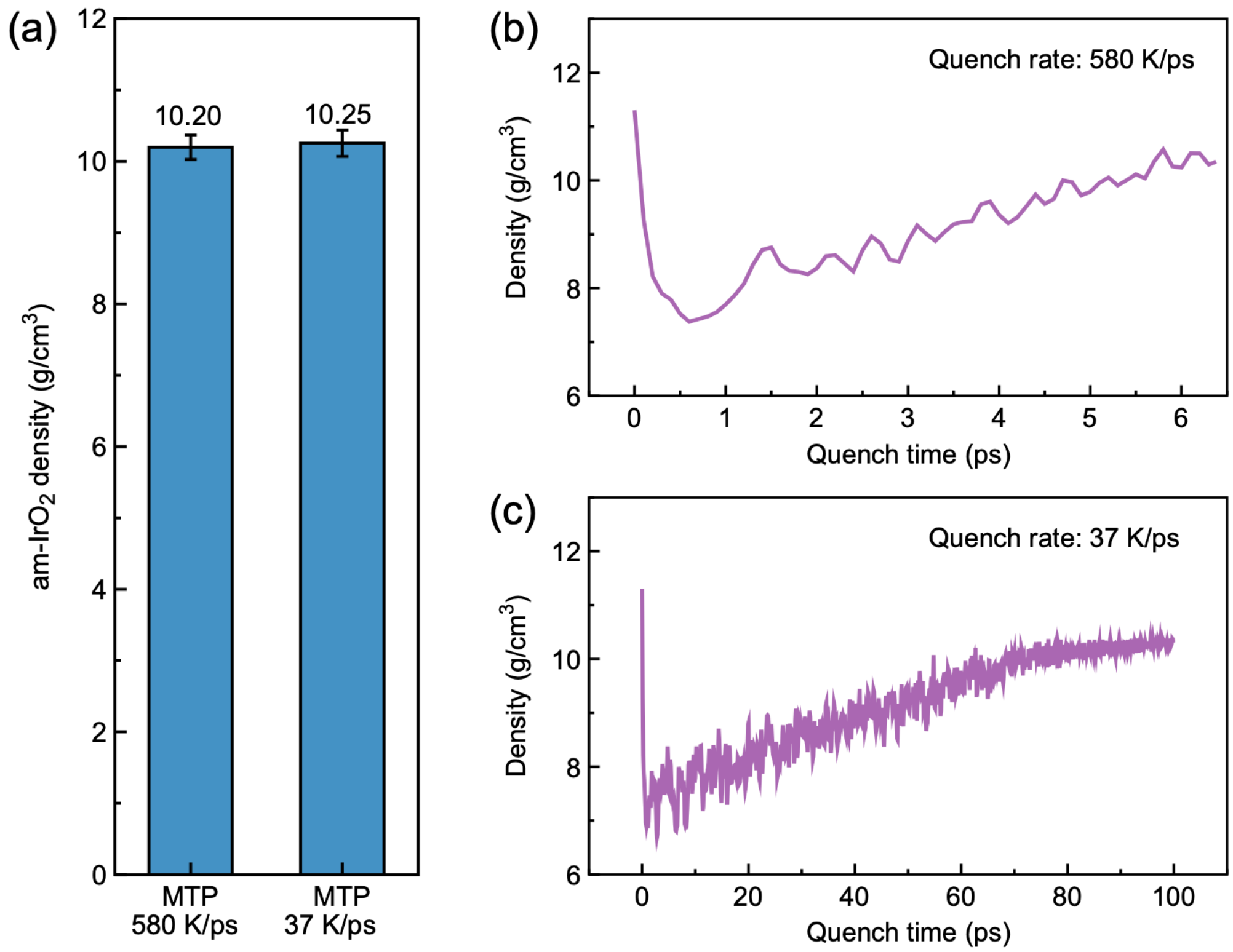

**Supplementary Fig. 2 | Effect of quench rate on MTP-generated am-$IrO_2$ density.** (a) Final density of am-$IrO_2$ generated by MTP at quench rates of 580 and 37 K/ps. Each bar represents the average density from five independent simulations, and the black error bars indicate the corresponding standard deviation. (b, c) Time evolution of the system density during quenching at 580 and 37 K/ps, respectively.

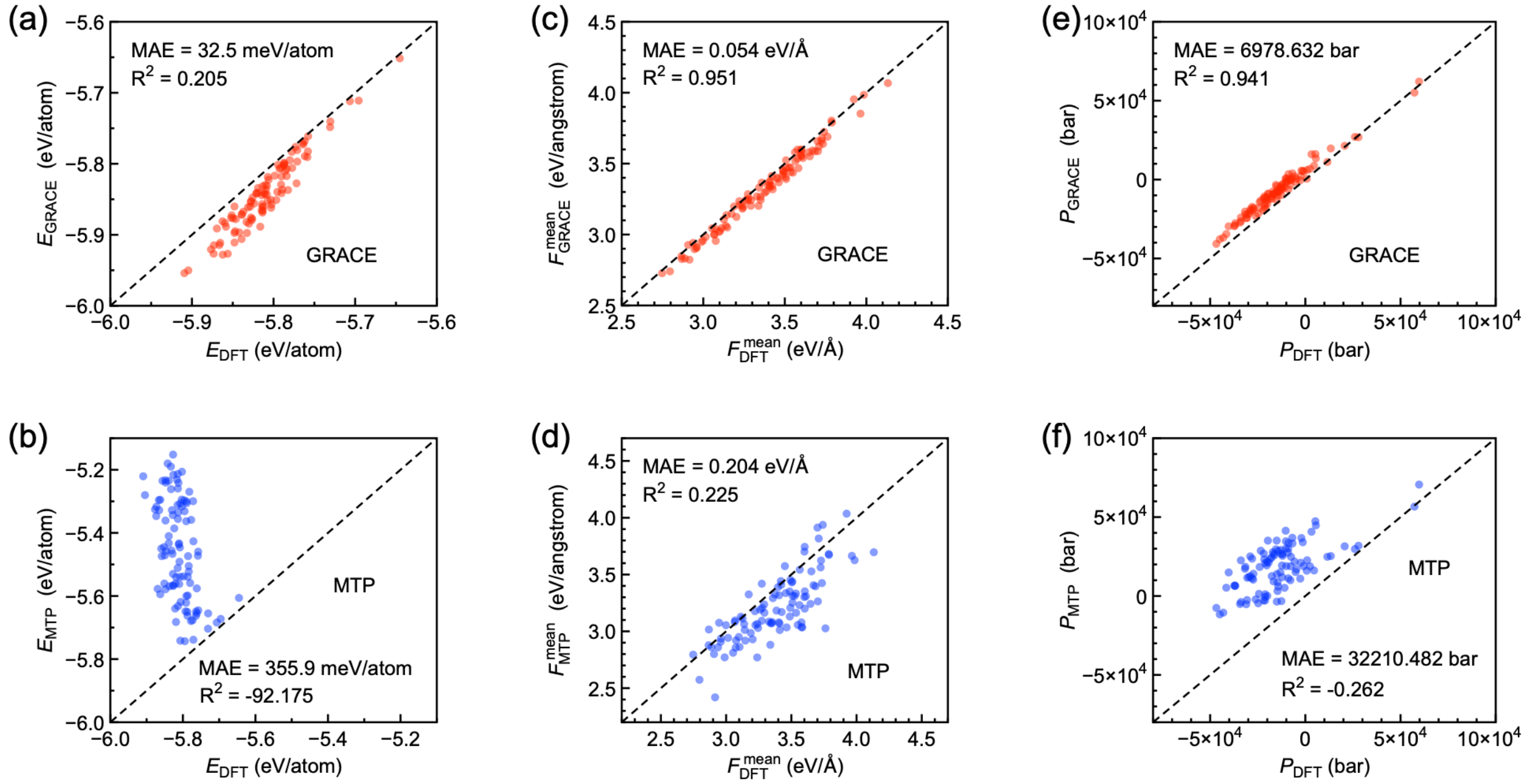

**Supplementary Fig. 3 | GRACE and MTP predictions compared with DFT references.** Parity plots compare predictions from GRACE (a, c, e) and MTP (b, d, f) with DFT reference values for (a, b) energy, (c, d) mean atomic force and (e, f) virial pressure.

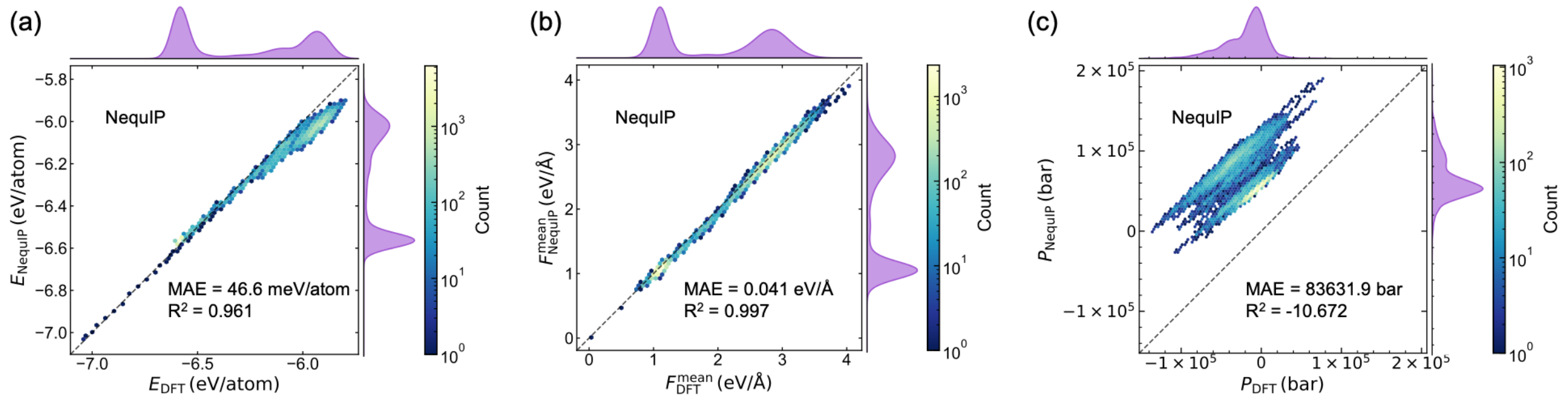

**Supplementary Fig. 4 | NequIP predictions validated against AIMD reference data.** Parity plots compare NequIP predictions with AIMD reference values for configurations sampled during AIMD: (a) energy, (b) mean atomic force and (c) virial pressure. Dashed black lines indicate one-to-one agreement, color maps show the number of sampled configurations in each bin and purple marginal distributions show the corresponding reference and predicted value distributions.

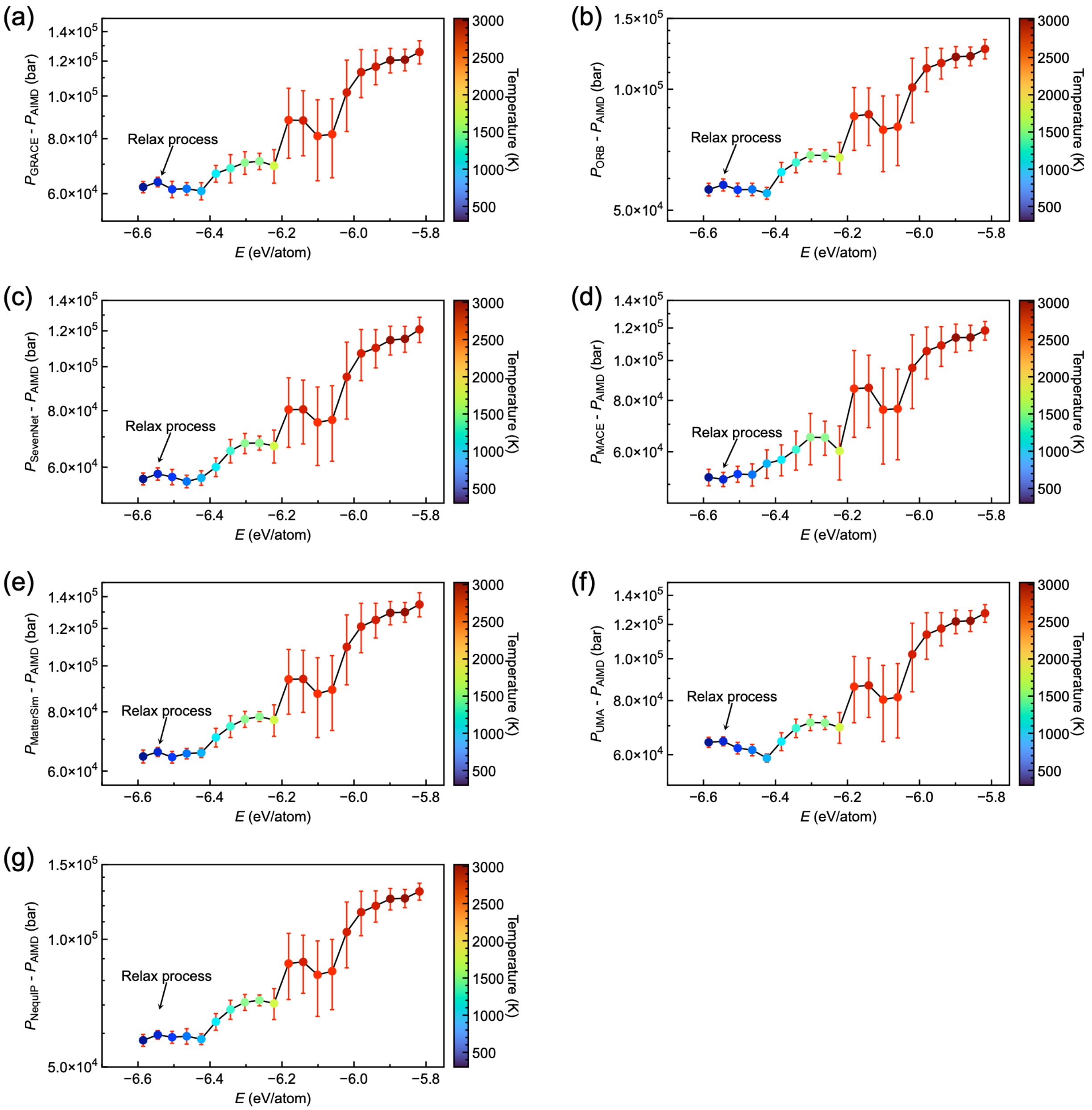

**Supplementary Fig. 5 | Temperature-dependent virial pressure deviations of uMLIPs from AIMD.** Virial pressure deviation, $P_{\mathrm{uMLIP}}$ - $P_{\mathrm{AIMD}}$, is plotted as a function of energy for (a) GRACE, (b) ORB, (c) SevenNet, (d) MACE, (e) MatterSim, (f) UMA and (g) NequIP. Data points and error bars represent the mean and standard deviation within each energy bin, respectively. Point color denotes temperature. Arrows mark the relaxation process.

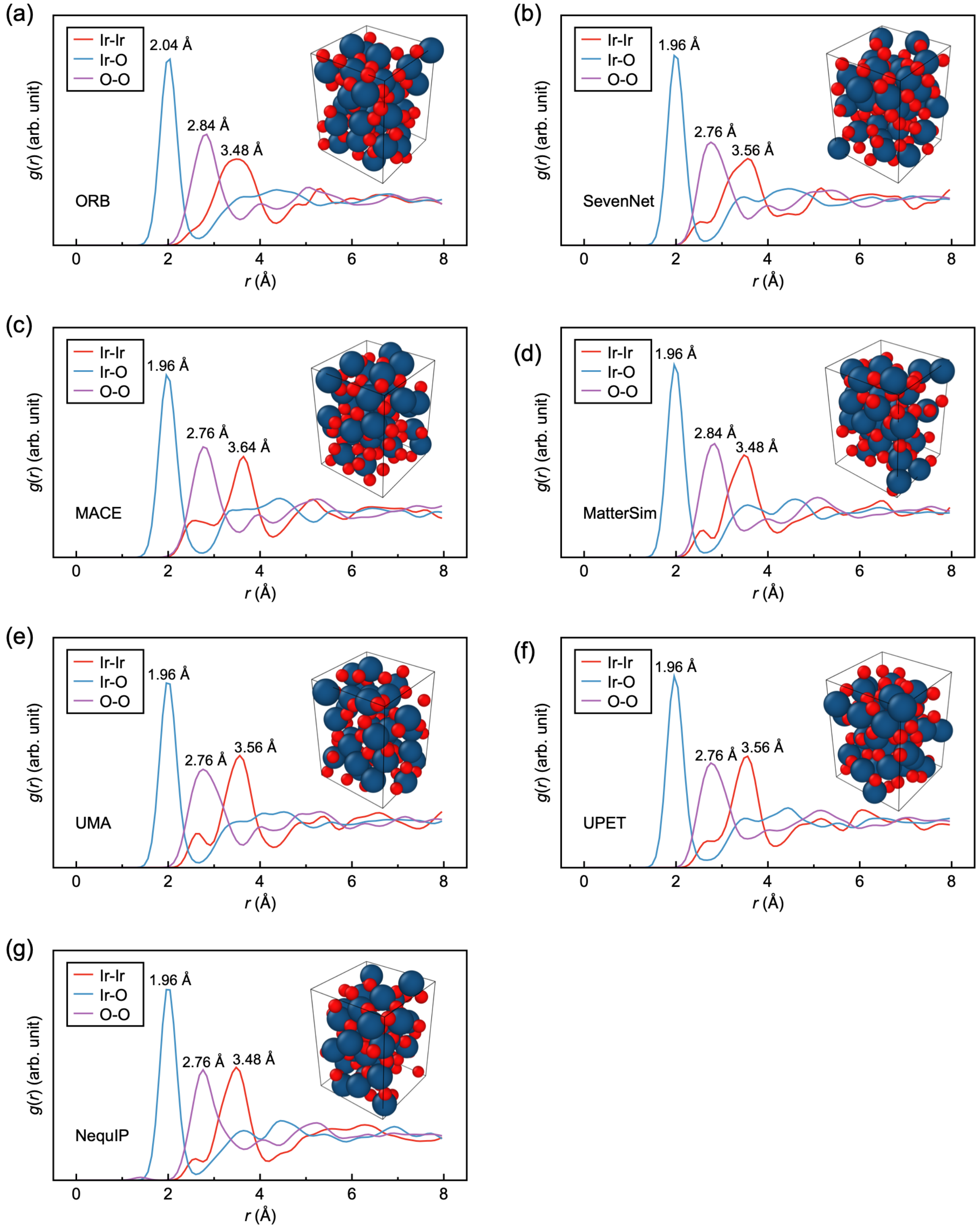

**Supplementary Fig. 6 | Structural characterization of am-$IrO_2$ generated with the revised protocol.** Radial distribution functions (RDFs), *g(r)*, for amorphous $IrO_2$ generated using the revised protocol with (a) ORB, (b) SevenNet, (c) MACE, (d) MatterSim, (e) UMA, (f) UPET and (g) NequIP. The inset shows the atomic configurations, with blue and red spheres denoting Ir and O atoms, respectively.